\documentclass[11pt,a4paper]{article}
\pdfoutput=1
\usepackage[british]{babel} 
\usepackage{a4wide}
\usepackage[pdftex]{graphicx} 
\usepackage[T1]{fontenc}
\usepackage[utf8]{inputenc}
\usepackage{varioref}\labelformat{equation}{(#1)}
\usepackage{fourier,charter} 
\usepackage[round]{natbib}
\usepackage{enumerate,mathrsfs}

\usepackage{amsfonts}
\newcommand{\anymode}[1]{\ifmmode{#1}\else\mbox{$#1$}\fi}

\newcommand{\NB}[1]{\texttt{/\textsuperscript{NB}}%
          \marginpar{\raggedright\sl\small#1\hfill}}	  

\long\def\Nota#1{\footnote{#1}\kern-0.2em\NB{cfr Nota$^{(\thefootnote)}$}}
\def\linea#1{\ifhmode\hfill\break\fi\hbox to \hsize{#1}}

\newcommand{\inv}{^{-1}}

\def\vc{v\kern -0.1em .c.\relax}

\newcommand{\cor}[2][]{
   \ensuremath{\textrm{cor}_{#1}\!\left\{\displaystyle{#2}\right\}}}

\newcommand{\cov}[2][]{
   \ensuremath{\textrm{cov}_{#1}\!\left\{\displaystyle{#2}\right\}}}
\newcommand{\dfrac}[2]{\displaystyle{\frac{#1}{#2}}}

\newcommand{\equald}{\stackrel{d}{=}}
\newcommand{\E}[2][]{
   \ensuremath{\mathbb{E}_{#1}\!\left\{\displaystyle{#2}\right\}}}

\newcommand{\half}{\mbox{$\textstyle \frac{1}{2}$}}   
\long\def\ignore#1{}

\newcommand{\indep}{\perp\kern-0.5em\perp}

\newcommand{\pr}[2][]{
   \ensuremath{\mathbb{P}_{#1}\!\left\{\displaystyle{#2}\right\}}}

\newcommand{\Real}{\mathbb{R}}

\newcommand{\T}{^{\top}}

\newtheorem{theorem}{Theorem}

\newtheorem{proposition}[theorem]{Proposition}
\newcommand{\N}{\mathrm{N}{}}

\renewcommand{\d}{\,\mathrm{d}\relax}

\title{\huge\bf 
   Selection models under generalized symmetry settings}
\author{{\Large Adelchi Azzalini}\\
   Dipartimento di Scienze Statistiche\\
   Università di Padova, Italia
   }
\date{\footnotesize 2nd April 2010}
\renewcommand{\phi}{\varphi}

\begin{document}

\maketitle

\begin{abstract}
  An active stream of literature has followed up the idea of
  skew-elliptical densities initiated by \citet{azza:capi:1999}. Their
  original formulation  was based on a general lemma which is however of  
  broader applicability than usually perceived. This note examines new
  directions of its use, and illustrates them with the construction 
  of some probability distributions falling outside the family
  of the so-called skew-symmetric densities.
\end{abstract}

\vspace{3ex} \emph{Key-words:} central symmetry, gamma distribution,
probability integral transform, skew-normal distribution, skew-symmetric
distributions, symmetric functions, symmetry.

\vspace{6ex}
\section{Background and aims} \label{s:intro}
A currently active stream of literature deals with $d$-dimensional continuous
probability distributions such that their density function can be written in
the form
\begin{equation}  \label{e:lemma1-pdf}
     f(z) = 2\,f_0(z)\,G\{w(z)\}, \qquad z\in\Real^d,
\end{equation}  
where $f_0$, called the `base function' in this paper, is a density function
satisfying some form of symmetry condition, and $G$ and $w$ are functions
whose requirements will be recalled shortly.  The more prominent
representative of this formulation is the skew-normal distribution whose
density function at $z$ is
\begin{equation}  \label{e:sn-pdf}
   2\:\phi_d(z;\Omega)\:\Phi(\alpha\T z) 
\end{equation} 
where $\phi_d$ denotes the $d$-dimensional normal density $\N_d(0,\Omega)$ and
$\Phi$ is the scalar standard normal distribution function 
\citep{azza:dval:1996}.
In the general formulation $\Omega$ is a covariance matrix, but for the 
present purposes we can restrict ourselves to the case where $\Omega$ is 
a correlation matrix; $\alpha$ is a vector of parameters which regulate 
the skewness of the distribution.
Clearly, setting $f_0(z)=\phi_d(z;\Omega)$, $G=\Phi$, $w(z)=\alpha\T z$ in
\ref{e:lemma1-pdf} produces \ref{e:sn-pdf}.

In fact the chronological order of development of the two expressions above
was the opposite. Near the end of a paper dedicated to the properties of
distribution \ref{e:sn-pdf}, \citet{azza:capi:1999} delineated a more general
formulation, starting from the following result, whose proof is reproduced
here because of it is extremely simple, yet instructive.

\begin{proposition} \label{th:lemma1999}
Denote by $G$ the distribution function of  a continuous random variable 
whose density function is symmetric about 0 and by $Y=(Y_1,\ldots,Y_d)\T$ 
a continuous random  variable with density function $f_0$.
If  the real-valued transform $w(Y)$ has symmetric density about 0,
then \ref{e:lemma1-pdf} is a  $d$-dimensional density function.
\end{proposition}
\noindent
\emph{Proof.} 
If $X\sim G$, independent of $Y$, the distribution of $X-w(Y)$ is symmetric 
about 0, implying that
\begin{equation} \label{e:proof-1999}
  \half = \pr{X\leq w(Y)} = \E[Y]{\pr{X\leq w(Y)|Y}}
        = \int_{\Real^d} G\{w(y)\}\,f_0(y)\,\d{y} \,.  
\end{equation}  

The first formulation descending from the above proposition assumed $f_0$ to 
be an elliptically contoured density centred at 0 and $w(y)$ a linear function, 
leading to what was later called the family of skew-elliptical densities.
Although the ensuing discussion mentioned that $f_0$ does not need to be
elliptical, the actual development of \citet{azza:capi:1999} focused
on this case.   The idea of skew-elliptical distributions has been
followed up and expanded by a several authors, including \citet{bran:dey:2001}, 
\citet{gent:lope:2005}, and a number of contributors to the book edited by
\citet{genton:2004-SE}.  

In these developments, it emerged that many results
could hold replacing the assumption of elliptical distribution for $f_0$ by
the weaker assumption of central symmetry.  From \citet{serfling:2006-ess2},
recall that a random variable $Y$ is said to be centrally symmetric about $0$
if $-Y$ has the same distribution of $Y$; if the density $f_0$, say, of $Y$ 
exists, then $f_0(y)=f_0(-y)$.  A quite general formulation is expressed 
in the following result presented by \citet{azza:capi:2003}.

\begin{proposition}  \label{th:lemma2003}
Denote by $f_0(\cdot)$ the density function of a $d$-dimensional
continuous random variable  which is centrally symmetric about $0$,
and by $G$ a scalar distribution function such that $G(-x)=1-G(x)$
for all real $x$. If $w(z)$ is a function from $\Real^d$ to
$\Real$ such that $w(-z)=-w(z)$ for all $z\in\Real^d$, then
\ref{e:lemma1-pdf} is a  $d$-dimensional density function.
\end{proposition}
 
An essentially equivalent result has been obtained independently by
\citet{wangJ:boye:gent:2004-ss}, with the component $G\{w(y)\}$ in
\ref{e:lemma1-pdf} replaced by a function $\pi(y)$ which must satisfy the
conditions $0\le \pi(y) \le 1$ and $\pi(y)+\pi(-y) = 1$ for all $y\in\Real^d$.
The term skew-symmetric density has subsequently been adopted to denote this
set of densities, or equivalently \ref{e:lemma1-pdf}, when the base function
$f_0$ is centrally symmetric, to emphasize the broader settings which includes
those with elliptical density as as subset.

An important result concerning these distributions is the existence
of a stochastic representation for a random variable $Z$ of this
type given by 
\begin{equation} \label{e:represent-2003}
  Z = \cases{ Y & if $X\le w(Y)$,\cr
              -Y & otherwise, }
\end{equation}  
where $X$ and $Y$ are independent random variable with distribution function
$G$ and density function $f_0$, respectively, and the functions $G$, $f_0$ and
$w$ satisfy the conditions of Proposition~\ref{th:lemma2003}.  This
representation is important not only for random number generation, but also
for obtaining further theoretical conclusions.  An especially important
implication is a distributional invariance property stating that, for any
function $t(\cdot)$ from $\Real^d$ to $\Real^q$ such that $t(y)=t(-y)$ for all
$y$, then $t(Z)$ and $t(Y)$ have the same distribution, written as
\begin{equation} \label{e:invar-2003}
     t(Y) \equald t(Z)\,.
\end{equation}

Much work has been dedicated in recent years to distributions constructed via
Proposition~\ref{th:lemma2003}, or its counterpart based on the function
$\pi(x)$.  In addition to those already quoted, the review paper of
\citet{azzalini:2005} provides many other references, and since then the list
has increased substantially.

However, Proposition~\ref{th:lemma2003} is a special case of 
Proposition~\ref{th:lemma1999}. Its role can be viewed in identifying
some very simple  conditions which ensure the fulfilment of those  of 
Proposition~\ref{th:lemma1999}. The aim of the present note is to examine
situations which fall under the setting of Proposition~\ref{th:lemma1999}
but not of Proposition~\ref{th:lemma2003}. 
More specifically, we shall consider cases where  $f_0$ is not symmetric
about $0$, or  $w(\cdot)$  is not an odd function. As a side effect,
this exploration leads to a deeper understanding of what has been
developed so far in connection to Proposition~\ref{th:lemma2003},
and it produces  a more general formulation of representation 
\ref{e:represent-2003} and of the invariance property \ref{e:invar-2003}.

Before tackling the specific target of this article, we note that the
statement of Proposition~\ref{th:lemma1999}  is still valid under somewhat
weaker assumpions, as follows. We can relax the assumption about absolute
continuity of all distributions involved, and allow $G$ or the distribution 
of $w(Y)$ to be of discrete or of mixed type, provided the condition
$\pr{X- w(Y) \le 0}=\half$ in \ref{e:proof-1999} still holds. A sufficient
condition to meet this requirement is that at least one of the random 
variables $X$ and $w(Y)$ is continuous. 

In our development, we shall however work mostly with the original 
formulation of  Proposition~\ref{th:lemma1999}, with only
a few remarks related to the above weaker assumptions. 

\section{Some general facts} \label{s:general}

In the framework of Proposition~\ref{th:lemma1999}, the  very argument of 
its proof justifies  the  representation of a random variable $Z$  with the
stated distribution as 
\begin{equation} \label{e:represent-1999}
  Z =   Y  \quad\hbox{if~} X \le w(Y) 
\end{equation}  
where the condition $X\le w(Y)$ is satisfied with probability \half.
It is seen that the density of $Z$ is proportional to the integrand 
function of the last term in \ref{e:proof-1999}.
There is an obvious sample selection mechanism which underlies the 
transformation of the distribution $f_0$ into $f$.

A natural question is whether a representation similar to
\ref{e:represent-2003} can hold.  The question is of theoretical interest, but
also of practical relevance, since in random number generation the use of
\ref{e:represent-2003} in place of \ref{e:represent-1999} avoids the rejection
of half of the $Y$ samples.  Notice that the non-rejection of samples in
\ref{e:represent-2003} is achieved by exploiting the symmetry of the
distribution $f_0$.
 
It appears that some additional conditions must be required to provide
a solution to the above problem. One such set of conditions is as follows.
If there exists an invertible transformation $R(\cdot)$  such that,
for all $y\in\Real^d$,
\begin{equation} \label{e:R}
    f_0(y)=f_0[R(y)], \qquad |\det R'(y)|=1, \qquad w[R(y)]= -w(y)\,,
\end{equation}  
 where $R'(y)$ denotes the Jacobian matrix of the 
partial derivatives, then 
\begin{equation} \label{e:represent-1999-R}
     Z = \cases{ Y & if $X \le w(Y)$,\cr
                R\inv(Y) & otherwise, }
\end{equation}  
has distribution \ref{e:lemma1-pdf}. In fact the density function of $Z$
at $z$ is 
\begin{eqnarray*}
  f(z) &=&  f_0(z)\,G\{w(z)\} + f_0(R(z))\:|\det R'(y)|\:[1-G\{w(R(z))\}] \\
       &=&  f_0(z)\,G\{w(z)\} + f_0(z) \:[1- G\{-w(z)\}] \\
       &=& 2\,f_0(z)\,G\{w(z)\} 
\end{eqnarray*}
using \ref{e:R} and $G(-x)=1-G(x)$.  
Under the weaker assumptions indicated at the end of Section~\ref{s:intro},
the latter equality may not hold for all $x$. However this may affect only 
a number of $x$ values  at most countable, and the density function $f(\cdot)$ 
can be replaced by a regularized version without affecting the distribution.

According to the first two conditions in \ref{e:R}, the density $f_0$ is 
required to behave according to a ``generalized symmetry'' with respect to 
$R(\cdot)$, that is $R\inv(Y)$  must have the same density  $f_0$  of $Y$. 
In addition $w(\cdot)$ must be an odd function in this
generalized sense, as required by the third condition \ref{e:R}.  
In Proposition~\ref{th:lemma2003}, in \ref{e:represent-2003} and in 
\ref{e:invar-2003}, the transformation function is $R(z)=-z=R\inv(z)$.

If in \ref{e:represent-1999} we reverse the sign of the inequality, this
generates the dual variable of $Z$, having density function
\[  2\,f_0(y) [1- G\{w(y)\}]=  2\,f_0(y) G\{-w(y)\}= 2\,f_0(y) G\{(w(R(y))\} \]
where the last equality makes use of the third condition \ref{e:R}.

For the analogous of the invariance property \ref{e:invar-2003},
consider a transformation $t(\cdot)$ from $\Real^d$ to $\Real^q$ which is
even in the adopted generalized sense, that is
\begin{equation} \label{e:t()}
     t(z) = t(R\inv(z)) 
\end{equation}  
for all $z\in\Real^d$. From representation \ref{e:represent-1999-R},
it is  then immediate that \ref{e:invar-2003} holds.

The function $R(\cdot)$ which satisfy \ref{e:R} does not need to be unique.
Typically, if $R$ is one such function, $R\inv$ is another one.  In some
cases, there are more than two functions $R$ which satisfy \ref{e:R}, as we
shall see in Section\,\ref{s:w-non-odd}.  \emph{Vice versa}, in other
cases $R\inv=R$, for instance when $R(z)=-z$, and we effectively have a single
transformation.
 
In the one-dimensional case a general way to construct a random variable with
symmetric distribution is via its integral transform. Specifically, if $Y$ has
distribution function $F_0$, then $F_0(Y)-\half$ is uniformly distributed on
$(-\half, \half)$ and application of Proposition~\ref{th:lemma1999} provides
the following conclusion.

\begin{proposition} \label{th:general-1d}
  Assume that $f_0$ is a probability density function with distribution
  function $F_0$ on $S_0\subset\Real$, $G$ is a distribution function over
  $\Real$ which assigns a  probability distribution symmetric about 0 and 
  it is continuous except possibly at 0, and $w_1$ is an odd function on 
  $(-\half,\half)$. Then
  \begin{equation} \label{e:case-1d}
    2\,f_0(x)\:G\{w_1[F_0(x)-\half]\}, \qquad 
   2\,f_0(x)\:G\{w_1[\half -F_0(x)]\}, \qquad(x\in S_0\subset\Real),
\end{equation}    
are density functions on $S_0$.
\end{proposition}
 
The same result could however have been equally obtained from
Proposition~\ref{th:lemma2003}, although via a slightly more involved
argument, as follows. Start by taking the base density function to be the
uniform density in $(-\half, \half)$, $w(z)=w_1(z)$, and then shift the
distribution to the interval $(0,1)$; we arrive at the density
$2\,G\{w_1(x-\half)\}$. 
Finally, transform this distribution via $F_0\inv(\cdot)$, to produce the 
first expression in \ref{e:case-1d}. The second expression is obtained
similarly using $w(z)=w_1(-z)$.
 
For $d>1$, the probability integral transform  is not so easily tractable
as for $d=1$. A manageable case occurs with $d=2$ and independent marginal 
components. If $Y_1$ and $Y_2$ are independent variables with
distribution functions $F_1$ and $F_2$, respectively, then a simple
computation gives
\[
  p(t)= \pr{F_1(Y_1)\,F_2(Y_2) \le t} = t\,(1-\log t) \,,
 \qquad (0<t<1)\,.
\]
Recall that $p[F_1(Y_1)\,F_2(Y_2)]$ is uniformly distributed in
$(0,1)$. If $f_1=F_1'$, $f_2=F_2'$ and $w_1$ is an odd function,
then from Proposition~\ref{th:lemma1999}
\[
  2\,f_1(y_1)\,f_2(y_2)\:
  G\left\{w_1\left(p[F_1(y_1) F_2(y_2)]-\half\right)\right\}
\] 
is a proper probability density on the support given by the Cartesian product
of the support sets of $F_1$ and $F_2$.

This mechanism is of general validity for $d=2$, but it has the disadvantage
that the base function $f_1(y_1)f_2(y_2)$ does not allow for dependence
between the marginal components. In addition, in many cases, the distributions
so produced are not very convenient to work with from a mathematical
viewpoint.  For these reasons, in the following section we examine other
mechanisms, which are simpler to handle. Moreover they are more directly
related to the formulation expressed by \ref{e:R}--\ref{e:t()}; hence they
better serve the purpose of illustrating the underlying mechanism.

\section{Some specific constructions}  \label{s:examples}

We shall now examine a few specific distributions to illustrate 
the formulation of the previous section.  Given this aim, a complete 
study of the properties of these distributions is not attempted here.
To ease exposition, we shall concentrate on the case $d=2$.

\subsection{Cases with  $w(y)$ non-odd}  \label{s:w-non-odd}

\begin{figure} 
   \centering
   \includegraphics[width=0.48\textwidth]{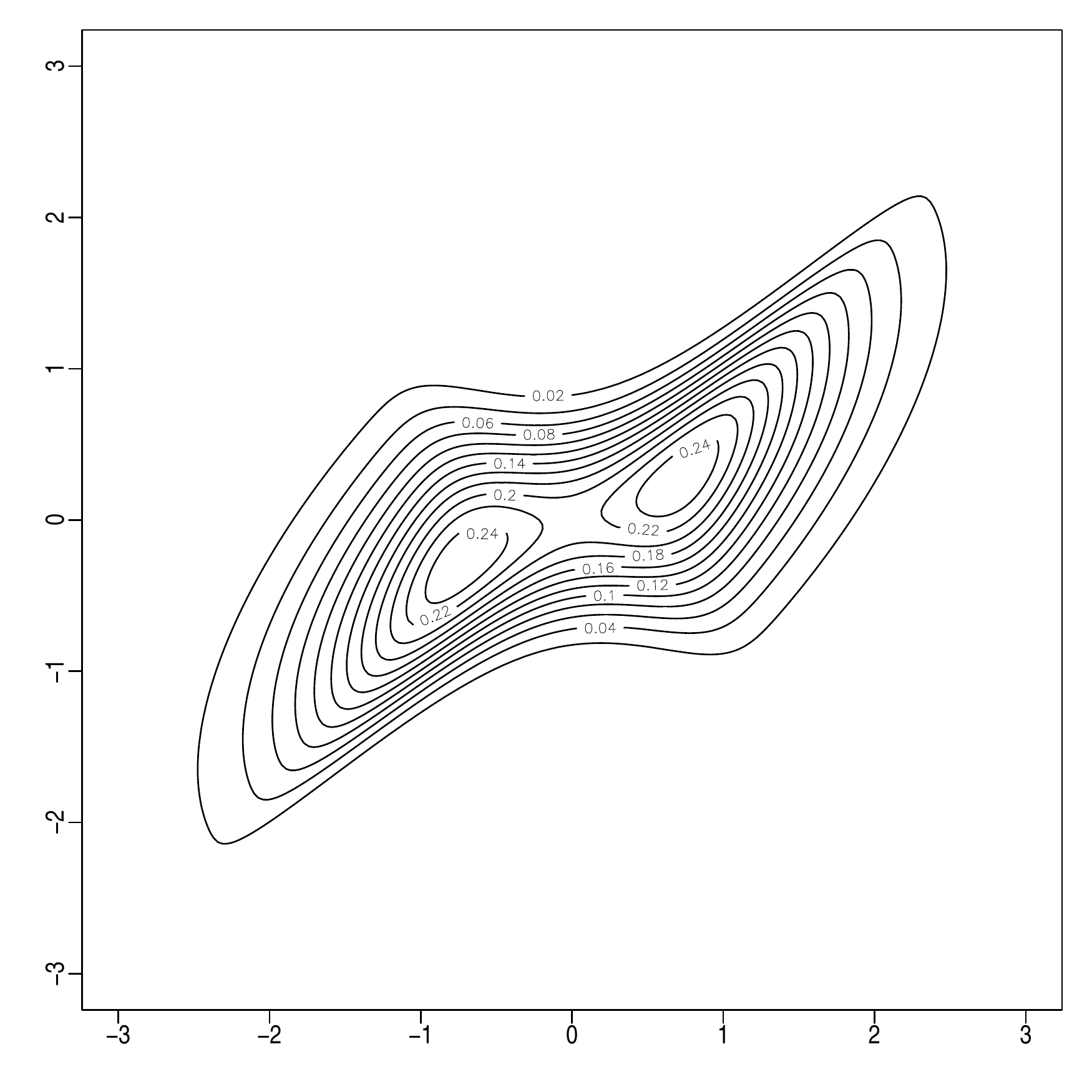} \hfil
   \includegraphics[width=0.48\textwidth]{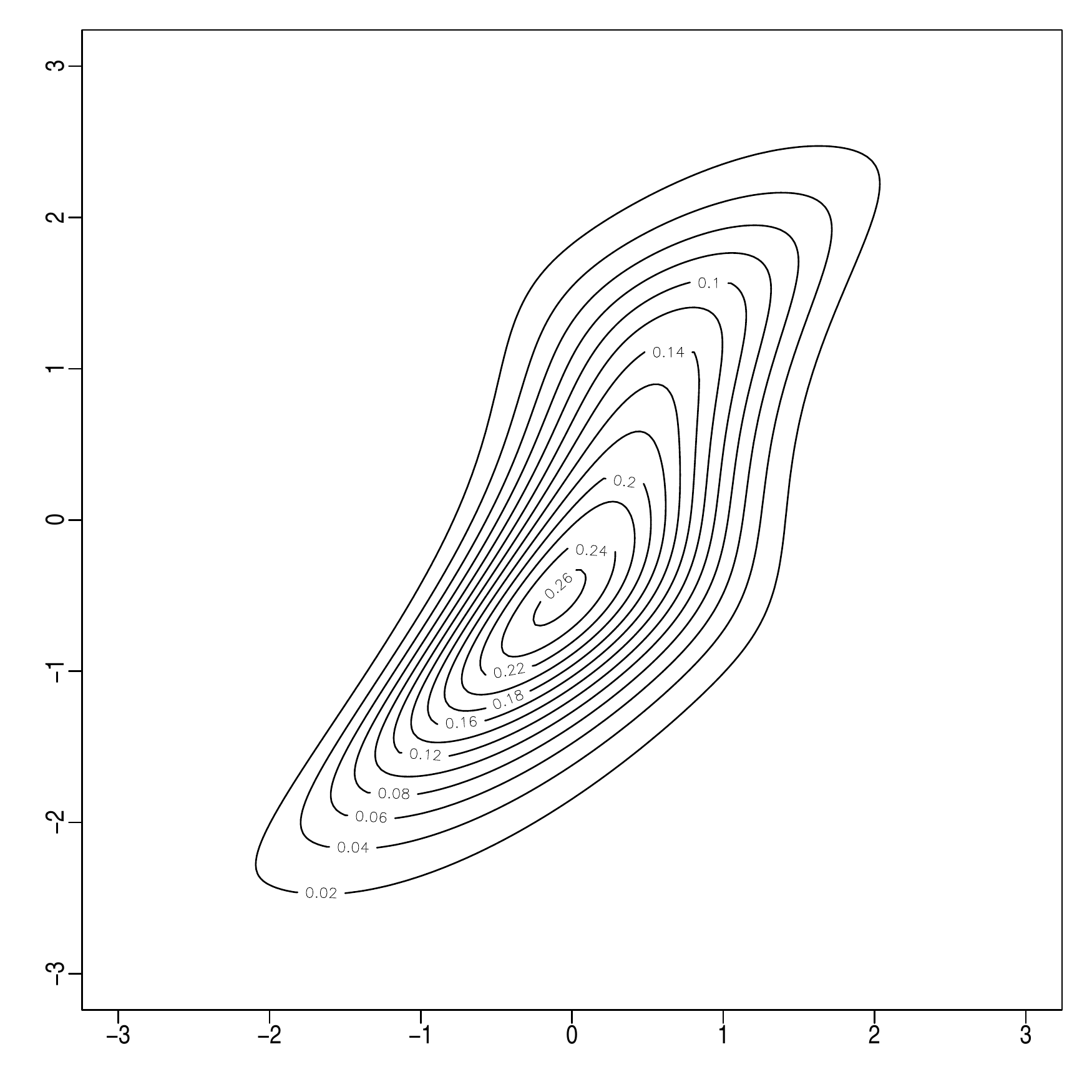}
   \caption{\sl Contour level plot of density function \ref{e:sn-diff-sq} when
     $\alpha=2$ and $\rho=2/3$ (left panel) and similar  density when the
     last factor is $\Phi\{z_1-z_2-(z_1^2-z_2^2)\}$ (right panel)}
   \label{fig:sn-diff}
\end{figure}

To start with a simple example, consider the probability distribution on
$\Real^2$ whose density function at $z=(z_1,z_2)\T$ is
\begin{equation} \label{e:sn-diff-sq}
      f(z) = 2\:\phi_2(z;\Omega)\,\Phi\{\alpha(z_1^2-z_2^2)\}
\end{equation}  
where $\Omega$ is a $2\times2$ correlation matrix with off-diagonal term
$\rho$, and $\alpha$ is a real parameter. This $f(z)$ bears a superficial
resemblance to the skew-normal density \ref{e:sn-pdf} for $d=2$, but with 
the noticeable difference from the skew-normal distribution that
\ref{e:sn-diff-sq} is not skew at all; in fact it is centrally symmetric.  The
effect of modification of the normal density when $\alpha=2$ and $\rho=2/3$ is
shown graphically in the left panel of Figure~\ref{fig:sn-diff}.  This density
is bimodal, but other choices of $\alpha$ produce a unimodal density.

The fact that the appropriate normalization constant in \ref{e:sn-diff-sq} 
is~2 cannot however follow from Proposition~\ref{th:lemma2003} which requires 
an odd function $w$, while $w(z)=\alpha(z_1^2-z_2^2)$ is even.
Equivalently, $\pi(z)=\Phi\{\alpha(z_1^2-z_2^2)\}$ does not satisfy the
condition $\pi(z)+\pi(-z)=1$ required in the formulation of
\citet{wangJ:boye:gent:2004-ss}.
However, if $Y=(Y_1,Y_2)\T\sim \N_2(0,\Omega)$, it is true that
$\alpha(Y_1^2-Y_2^2)$ has a symmetric distribution about 0, and so
Proposition~\ref{th:lemma1999} apply to conclude that \ref{e:sn-diff-sq}
integrates to $1$.  In this respect, it would be irrelevant to replace $\Phi$
in \ref{e:sn-diff-sq} by some other distribution $G$ with $G'$ even.

Much more general forms than \ref{e:sn-diff-sq} can however be handled,
taking into account the following statement, whose proof is trivial and
omitted. 

\begin{proposition} \label{th:f0*w2}
If $(X,Y)$ is a  random variable on $\Real^2$ with density function 
$h(x,y)$ which is a symmetric function of the variables, that is $h(x,y)=h(y,x)$,
and $w_2(x,y)$ is a real-valued function such that $w_2(y,x)=-w_2(x,y)$, then 
$w_2(X,Y)$ has symmetric  distribution around $0$.
\end{proposition}  
A simple example of function $w_2$ which fulfils the above requirement is  
\begin{equation}  \label{e:w2-poly}
    w(z)=w_2(z_1,z_2)= \alpha_1(z_1-z_2)+\cdots+\alpha_m(z_1^m-z_2^m)
\end{equation} 
where $z=(z_1,z_2)\T\in\Real^2$ and $m$ is some natural number.  
The right-side panel of Figure~\ref{fig:sn-diff}
shows the contour level plot of the density  obtained when the argument of
$\Phi$ in \ref{e:sn-diff-sq} is replaced by \ref{e:w2-poly} with $m=2$,
$\alpha_1=1$, $\alpha_2=-1$, and $\rho=2/3$ as before.  Notice that in
general \ref{e:w2-poly} is neither odd nor even, if the coefficients
$\alpha_1,\dots,\alpha_m$ are unrestricted.  It is even if all $\alpha_j$'s of
odd order are 0, and it is odd in the dual case with only coefficients of odd
order, we are back to the setting of Proposition~\ref{th:lemma2003}.  
Combining Proposition~\ref{th:lemma1999} with Proposition~\ref{th:f0*w2}, 
we can state the following corollary.
\begin{proposition} \label{th:lemma1-w2}
If $f_0(z)$ is a density function centrally symmetric over $\Real^2$ and
$f_0$ is a symmetric function of the components of $z$,
$w_2$ is as in Proposition~\ref{th:f0*w2} and $G$ is a continuous 
distribution function over $\Real$ which assigns a  probability distribution
symmetric about 0 and it is continuous except possibly at 0,  then
\begin{equation}  \label{e:lemma1-w2}
  2\,f_0(z)\,G\{w_2(z_1,z_2)\}\,, \qquad z=(z_1,z_2)\T\in\Real^2,
\end{equation}
is a density function over $\Real^2$. 
\end{proposition}  

For simplicity, we now restrict ourselves to the case where
 $f_0(z)$ in \ref{e:lemma1-w2} is $\phi_2(z;\Omega)$ and $w_2$ is as in
Proposition~\ref{th:f0*w2}, but the essence of the conclusions remains true
for other centrally symmetric densities $f_0$, symmetric of the arguments.  
The stochastic representation
\ref{e:represent-1999-R} for a random variable $Z=(Z_1,Z_2)\T$ with such
density function holds by choosing a transformation which swaps the
two variables, that is
\begin{equation}   \label{e:R0}
    R(y) = R_0\:y\,, \qquad R_0=\pmatrix{0 & 1\cr 1 & 0} = R_0\inv\,,
\end{equation}  
and consequently also \ref{e:t()} applies for suitable $t$'s. Among the many
implications of this fact, the Mahalanobis distance $Z\T\Omega\inv Z$ is
$\chi^2_2$.  
Another consequence is that, since $t(y)=y_1 y_2=y_2\,y_1=t(R_0 y)$, then
$\E{Z_1\,Z_2}=\E{Y_1\,Y_2}=\rho$.  When central symmetry of $f(z)$ holds, for
instance when \ref{e:w2-poly} is even, this implies that $\E{Z_1}=\E{Z_2}=0$,
and then $\cov{Z_1,Z_2}=\rho$.

\begin{figure} 
   \centering
   \includegraphics[width=0.48\textwidth]{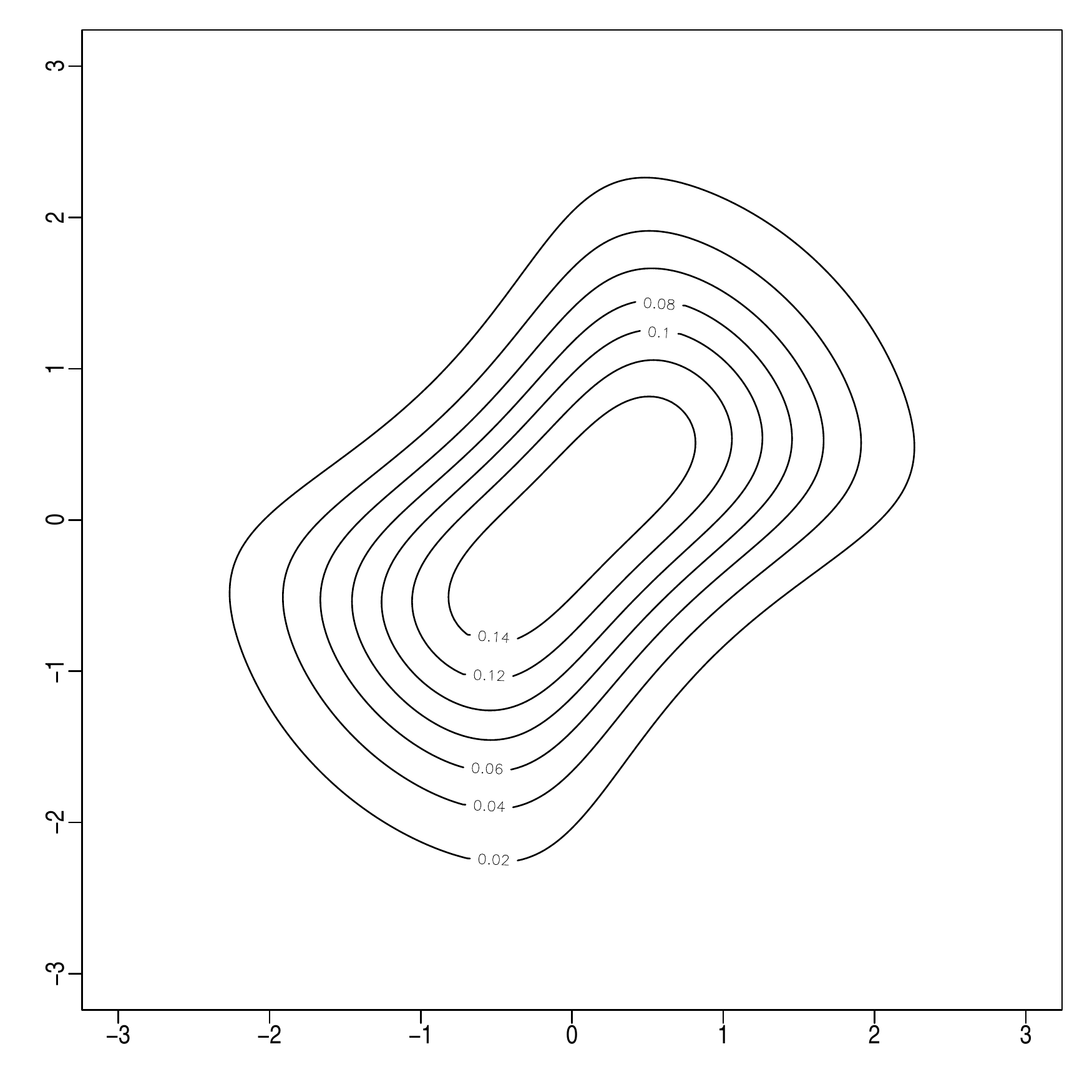} \hfil
   \includegraphics[width=0.48\textwidth]{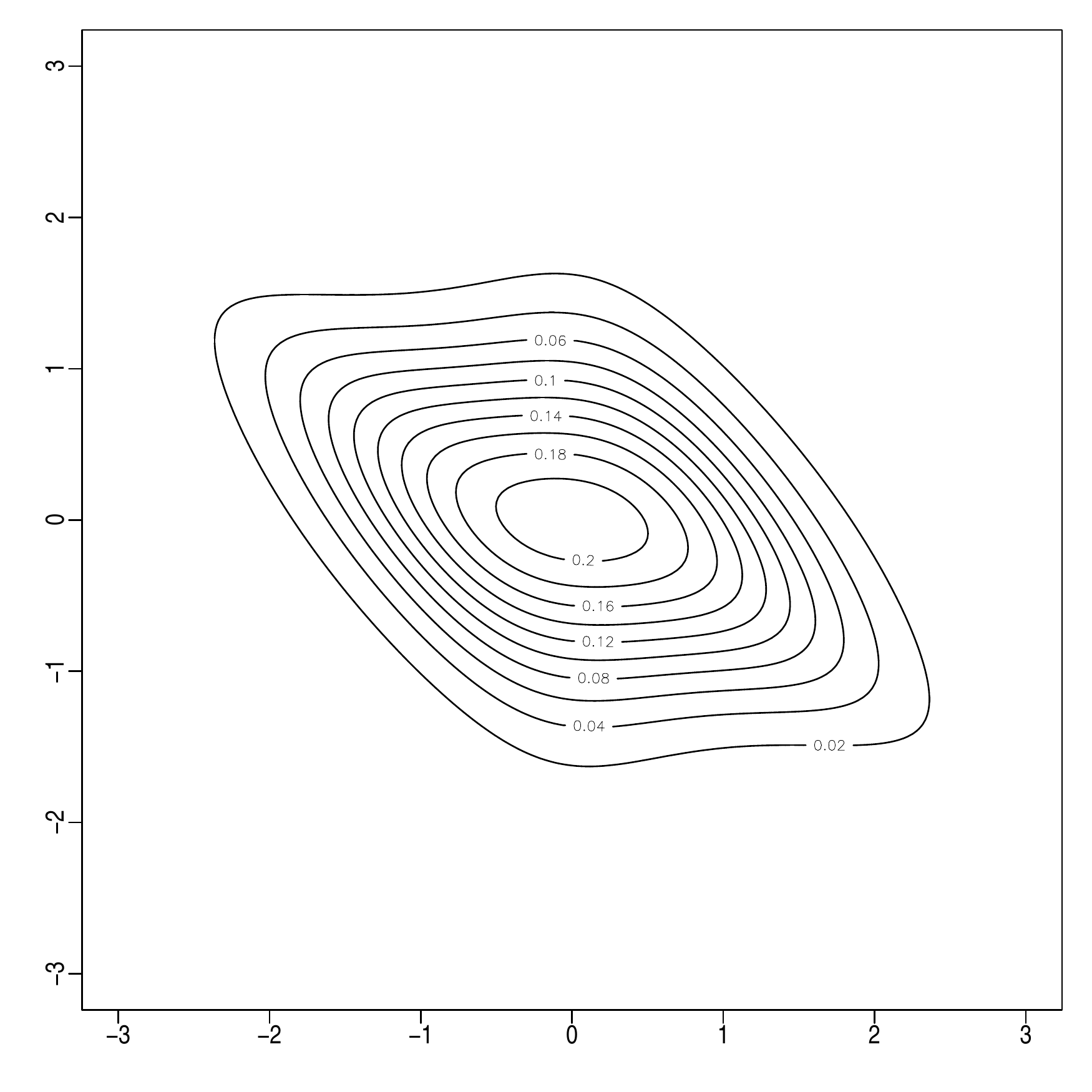}
   \caption{\sl Contour level plot of density function \ref{e:sn-product} when
     $\alpha=\sqrt{\pi/2}$ (left panel) and  density function given by the  
     first expression in \ref{e:sn-product-rho} when $\rho=-2/3$ and  
     $\alpha=\sqrt{\pi/2}$ (right panel)}
   \label{fig:sn-prod}
\end{figure}

We now move to a different example, and consider the bivariate distribution 
studied by \citet{arno:cast:sara:2002} whose density function at
$z=(z_1,z_2)\T\in\Real^2$ is
\begin{equation} \label{e:sn-product}
    f(z)= 2 \, \phi_2(z;I_2)\,\Phi(\alpha z_1\, z_2), \qquad            
\end{equation}
which is centrally symmetric too, and it enjoys several interesting
properties.  If $Z=(Z_1,Z_2)\T$ is a random variable with this distribution,
then each of $Z_1$, $Z_2$ is marginally standard normal, and the distribution
of each component conditional on the other one is univariate skew-normal
with parameter $\alpha$, that is of type \ref{e:sn-pdf} with $d=1$. Other
properties are presented in the quoted paper, including an expression for
$\cor{Z_1, Z_2}$. A graphical representation of this density is shown in 
the left-side panel of Figure~\ref{fig:sn-prod} for the case
$\alpha=\sqrt{\pi/2}\approx 1.253$ which has been proved by
\citet{arno:cast:sara:2002} to be the highest possible value producing a
unimodal density; with larger $\alpha$ there are two modes.

From the viewpoint of present context, \ref{e:sn-product} has a function
$w(z)=\alpha z_1 z_2$ which is even, and again of
Proposition~\ref{th:lemma2003} does not apply. It is however true that $w(Y)$
has a symmetric distribution around $0$, if $Y\sim\N_2(0,\,I_2)$, and
Proposition~\ref{th:lemma1999} confirms that $2$ is the appropriate
normalization constant in \ref{e:sn-product}.  More importantly, this explains
that the normalizing $2$ factor required in \ref{e:sn-product} is not by
accident, and it connects the distribution of \citet{arno:cast:sara:2002} 
with an apparently unrelated type of construction.

To fulfil conditions \ref{e:R} we can 
choose $R(y)$ to be any of $R_j y$ for $j=1,\dots,4$ where
\begin{equation} \label{e:R1-4}
   R_1 = \pmatrix{0 & -1 \cr 1 & 0}, \quad
   R_2 = -R_1, \quad
   R_3 = \pmatrix{-1 & 0 \cr 0 & 1}, \quad
   R_4 = -R_3 \,.
\end{equation}
The first two of these matrices correspond to  $\pi/2$ and $-\pi/2$ rotation, 
respectively, hence $R_2=R_1\T=R_1\inv$ and $R_1^4=I_2$. 
This provides a stochastic representation for $Z$, and so a
mechanism for random number generation.  Using \ref{e:t()} with
$t(z)=z_1^2+z_2^2$, we can state that $Z_1^2+Z_2^2\sim\chi^2_2$.
Using $t(z)=(z_1\,z_2)^2$ we obtain that $\E{(Z_1\,Z_2)^2}=1$, which in
turns implies $\cor{Z_1^2, Z_2^2}=0$, recalling that $Z_1$ and $Z_2$ have
standardized marginals.

If in \ref{e:sn-product} we want to replace $I_2$ by a correlation matrix
$\Omega$ with off-diagonal element $\rho$, then $w(Y)=\alpha Y_1 Y_2$ does no
longer give rise to a symmetric distribution around 0 if $(Y_1, Y_2)\T\sim
N_2(0,\Omega)$; hence Proposition~\ref{th:lemma1999} does not apply.  The
symmetry condition is fulfilled instead by $w(Y)=\alpha Y_1(Y_2-\rho Y_1)$ and
by the dual function $w(Y)=\alpha Y_2(Y_1-\rho Y_2)$, since we have again the
product of independent normal variables with $0$ mean. It then follows that
\begin{equation} \label{e:sn-product-rho}
    2 \,\phi_2(z;\Omega)\:\Phi\{\alpha z_1(z_2-\rho z_1)\}\,,\qquad
    2 \,\phi_2(z;\Omega)\:\Phi\{\alpha z_2(z_1-\rho z_2)\}\
\end{equation}
are legitimate density functions on $\Real^2$. The right-side panel of
Figure~\ref{fig:sn-prod} displays the density corresponding to $\rho=-2/3$
and $\alpha=\sqrt{\pi/2}$ as before.

\begin{figure} 
   \centering\hfil
   \includegraphics[width=0.45\textwidth]{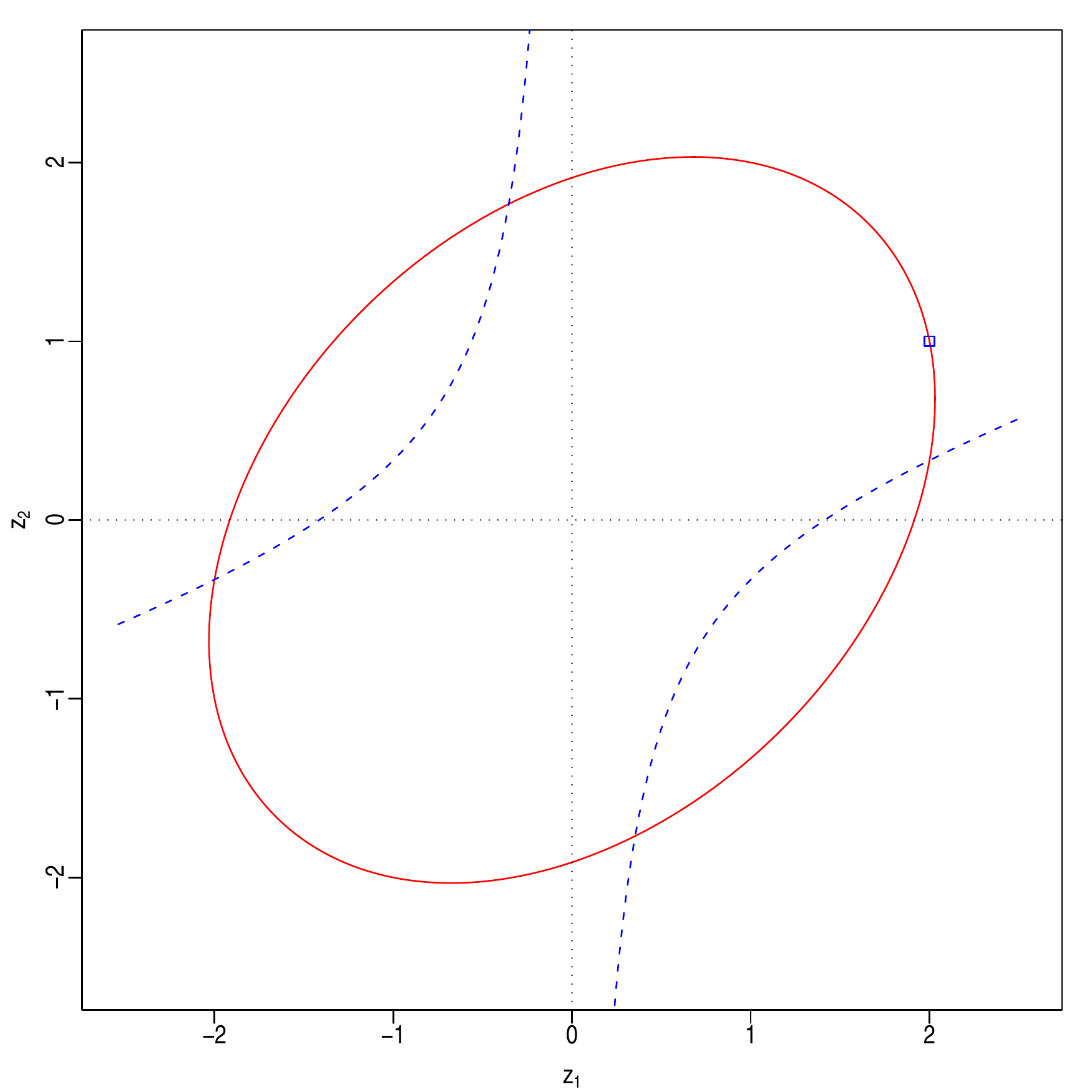} \hfil
   \includegraphics[width=0.45\textwidth]{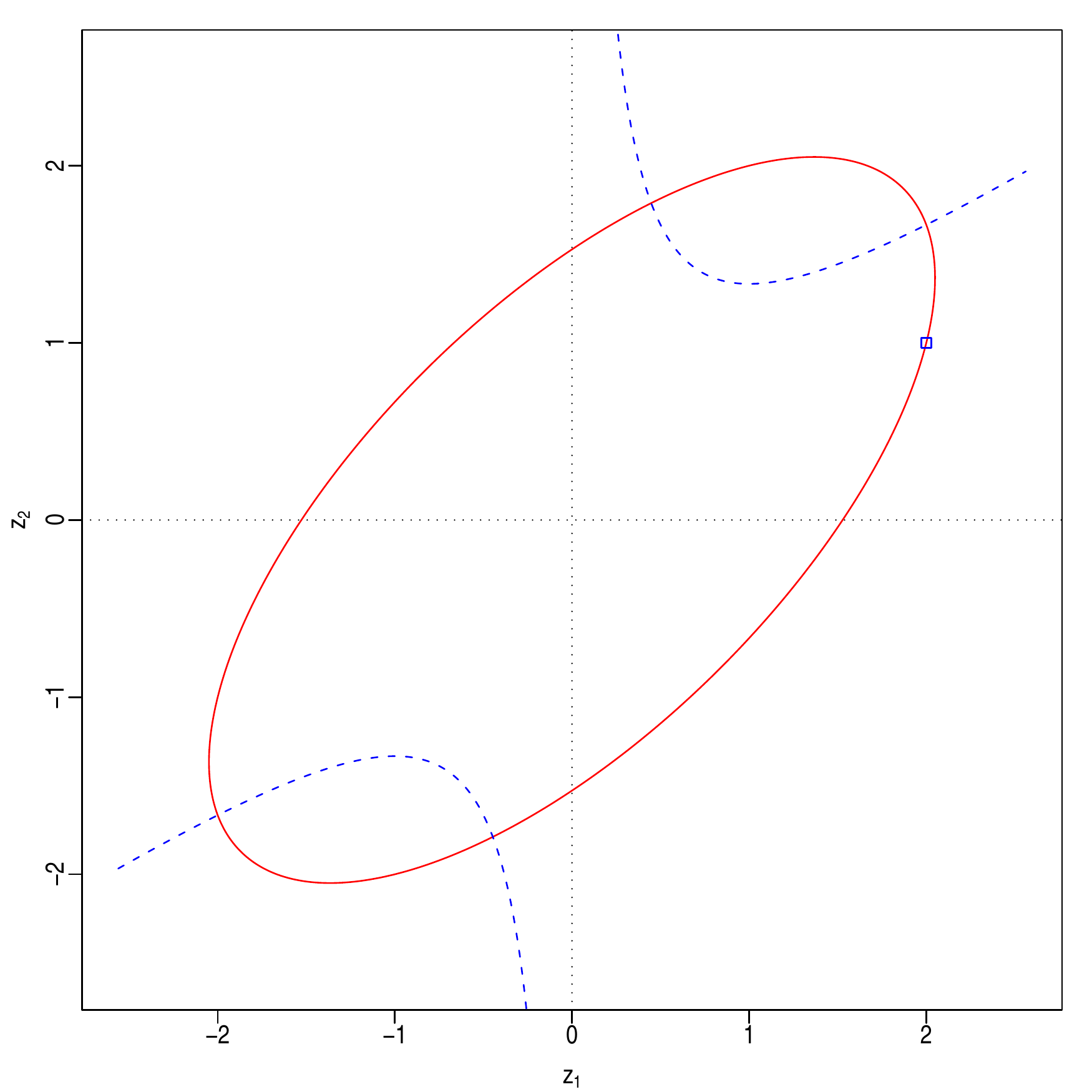} \hfil
   \caption{\sl Illustration of the possible choice of $R(z_0)$ for the first
     of the densities \ref{e:sn-product-rho} when $z_0=(2,1)\T$, $\alpha=1$
     and $\rho=1/3$ (left panel) or $\rho=2/3$ (right panel). The ellipse
     represents the locus of points with the same density of $z_0$; the dashed
     lines denote the loci of points with $w(z)$ equal to $-w(z_0)$}
   \label{fig:ellipse-w(z)}
\end{figure}

To search for a function $R(\cdot)$ fulfilling \ref{e:R}, we start by 
imposing the first and the third condition indicated by \ref{e:R}.
For each given point $z_0\in\Real^2$, it is required to find a point
which lies on the ellipse representing the locus of all points having the same
density of $z_0$ and it also lies on the locus of all points having $w(z)$
equal to $-w(z_0)$, that is to solve the equations
\[  D(z)= D_0, \qquad w(z)=-w_0 \]
where 
\[ 
  D(z)= z\T \Omega\inv z=\frac{1}{1-\rho^2}(z_1^2-2\rho z_1 z_2+z_2^2)\,,
  \qquad 
  w(z) = z_1 z_2-\rho z_1^2 
\]
and $D_0=D(z_0)$, $w_0=w(z_0)$. After simple algebra one arrives at a fourth
degree equation for $z_1$ with only even-order terms, or equivalently to the
quadratic equation
\[
   (1-\rho^2) u^2 - D_0(1-\rho^2) u + w_0^2 = 0 
\] 
where $u=z_1^2$. The roots of this equation lead to four values for
$z_1=\pm\sqrt{u}$ and four corresponding $z_2=(\rho z_1^2-w_0)/z_1$.
There is a clear similarity with the earlier case where $\rho=0$
and four transformations associated to matrices \ref{e:R1-4},
but now the four transformations $R(\cdot)$ are non-linear.
The construction is illustrated by Figure~\ref{fig:ellipse-w(z)} for the
given point $z_0=(2,1)\T$, $w(z)=z_1(z_2-\rho z_1)$ and $\rho$ equal 
either to $1/3$ or to $2/3$; the change of pattern in the two cases
depends on the sign of $w(z_0)$ as $\rho$ varies.  
 
The final step is to check whether the second condition in \ref{e:R} holds.
Some algebraic work nor reported here shows that $|\det R'|\not\equiv1$.
The implication is that, while \ref{e:sn-product-rho} are valid density
functions, a stochastic representation of type \ref{e:represent-1999-R}
does not hold for them. However \ref{e:represent-1999} still holds.
\subsection{Cases where $f_0$ is not a symmetric density}  \label{s:f0-asymm}

There are cases where $Y$ is a continuous random variable with density
function $f_0$ which is not symmetric about 0 but there still are functions
$w(Y)$ with symmetric density about 0, so that Proposition~\ref{th:lemma1999}
can be employed. Some constructions of this form have been sketched in
Section~\ref{s:general}.

If $d=2$, a simple but quite general formulation of this type can be obtained
when $f_0(\cdot)$ and $w(\cdot)$ satisfy the requirements for $h$ and
$w_2$ in Proposition~\ref{th:f0*w2}, respectively. 
In other words, we are asking that  
\[
    f_0(y)=f_0(R_0 y), \qquad w(R_0 y)= -w(y) \, 
\]
where $R_0$ is as in \ref{e:R0}. 
Under this setting, \ref{e:represent-1999-R} holds with $R(y)=R_0\,y$. 

A simple illustrative example of this formulation can be started by choosing
$f_0$ to be the product of two Gamma densities, such that each marginal
density is
\[
    f_1(x) = \frac{1}{\Gamma(\omega)} x^{\omega-1}\:e^{-x}  
\]
if $x>0$, and 0 otherwise, for some $\omega>0$. 
For $f_0(y)=f_1(y_1)\,f_1(y_2)$ where $y=(y_1,y_2)\T$, 
there is a natural line of reflection  given by the identity line. 
Among the many feasible functions $w$, we choose  the simple 
linear case $w(y)=\alpha(y_1-y_2)$ where $\alpha$ is an arbitrary parameter.
The density under consideration is then
\begin{equation}  \label{e:f(z)-gamma}
  f(z_1,z_2) = \frac{2}{\Gamma(\omega)^2} (z_1\,z_2)^{\omega-1}
          \exp(-z_1-z_2)\:G\{\alpha(z_1-z_2)\}
\end{equation}
for $z_1,z_2>0$, and $0$ otherwise. Here $G$ is as required in
Proposition~\ref{th:lemma1999} with possibly a discontinuity at $0$. 
If $Y=(Y_1,Y_2)\T$ has density $f_0$, then
$f(z_1,z_2)$ is the distribution of $Y$ conditionally on the event
$\{X<\alpha(Y_1-Y_2)\}$, where $X\sim G$ is independent of $Y$.

Various implications follow from \ref{e:invar-2003} for  a random variable
$Z=(Z_1,Z_2)\T$ having density function \ref{e:f(z)-gamma}. 
Some examples are
\[ 
   \E{Z_1 + Z_2} = \E{Y_1+Y_2} = 2\,\omega \,, \qquad
   \E{Z_1\:Z_2} = \E{Y_1\:Y_2} = \omega^2 \,, 
\]
irrespectively of $G$, since both  $t(y)=y_1+y_2$ and $t(y)=y_1\,y_2$ 
satisfy \ref{e:t()}.

To compute the marginal distributions of $Z_1$ and $Z_2$, assume for the
moment that $\alpha>0$, and rewrite the event $\{X<\alpha(Y_1-Y_2)\}$ as
$\{T<Y_1\}$ where $T=\alpha\inv\,X +Y_2$.  Denote by $F_T$ the distribution
function of $T$, which is the integral of the density
\begin{equation} \label{e:T-pdf}
    f_T(t) = \int_0^\infty f_1(y_2)\:g[\alpha(t-y_2)]\alpha\:\d{y_2} 
\end{equation}  
where $g=G'$. It is simple to obtain  the density functions of $Z_1$ 
and $Z_2$ which are
\begin{equation} \label{e:Z1-Z2-pdf}
   2\,f_1(z_1)\:F_T(z_1), \qquad  2\,f_1(z_2)\:\{1-F_T(z_2)\}, \qquad 
   (z_1>0,\ z_2>0), 
\end{equation}
respectively, taking into account that $\pr{T<Y_1}=\half$.
When $\alpha<0$, the density function of $T$ is still computed from
\ref{e:T-pdf}, with $\alpha$ replaced by its absolute value,
since the distribution of $\alpha\inv X$ does not depend on the sign
of $\alpha$, and the above expressions of the densities of $Z_1$ and $Z_2$
are exchanged. Another way of  looking at this aspect is to take into
account \ref{e:represent-1999-R} and to notice that reversing the sign
of $\alpha$ corresponds to swap the distribution of the two components.

The explicit computation of \ref{e:T-pdf} is feasible for a suitable
form of $g$. A convenient option is to set $g$ equal to the Laplace
density
\[
   g(x)= \half \:e^{-|x|}, \qquad (x\in\Real),
\]
leading to
\begin{eqnarray*}
  f_T(t) &=& \cases{ 
    \dfrac{\alpha}{2(1+\alpha)^\omega}\:e^{\alpha t} & if $t\le 0$, \cr
      \vphantom{\rule{0pt}{3ex}}
      \dfrac{\alpha}{2\,\Gamma(\omega)}
      \left(e^{-\alpha t} I_1 +e^{\alpha t}\:I_2\right)
     & otherwise,
    }
\end{eqnarray*}  
where, for integer $\omega$,
\begin{eqnarray*}
  I_1 &=& \int_0^t x^{\omega-1} e^{-x(1-\alpha)}\:\d{x}\\
      &=& \cases{
        \dfrac{t^\omega}{\omega} & if $\alpha=1$,\cr   
         \vphantom{\rule{0pt}{3ex}}
        \Gamma(\omega)\left\{\dfrac{1}{(1-\alpha)^\omega}
          -e^{-t(1-\alpha)} 
           \sum_{k=0}^{\omega-1} \dfrac{t^k}{k! \,(1-\alpha)^{\omega-k}}
           \right\}
                          & if $\alpha\not=1$,
      }\\
  I_2 &=&\int_t^\infty x^{\omega-1} e^{-x(1+\alpha)}\:\d{x}\\
      &=& \Gamma(\omega) e^{-t(1+\alpha)} 
           \sum_{k=0}^{\omega-1} \frac{t^k}{k! \,(1+\alpha)^{\omega-k}}\,.
\end{eqnarray*}  

It is possible to produce a general expression of the distribution function
$F_T$. However, for simplicity,  we now  restrict ourselves to the case
$\omega=1$, such that
\[
  F_T(t) = \cases{
           \dfrac{1}{2(1+\alpha)}e^{\alpha t} & if $t\le 0$, \cr
            \vphantom{\rule{0pt}{3ex}}
           1 -e^{-t}\left(\frac{3}{4}+ \frac{t}{2}\right) 
                                & if $t> 0, \ \alpha=1$,\cr
           1- \dfrac{\alpha^2}{\alpha^2-1} e^{-t} 
             + \dfrac{1}{2(\alpha-1)} e^{-\alpha t}  
                                & if $t> 0, \ \alpha\not=1$,
          }
\]
when $\alpha>0$. 
By inserting this expression of $F_T(t)$ in \ref{e:Z1-Z2-pdf}, we
obtain the marginal density functions of $Z_1$ and $Z_2$.
\begin{figure} 
   \centering
   \includegraphics[width=0.48\textwidth]{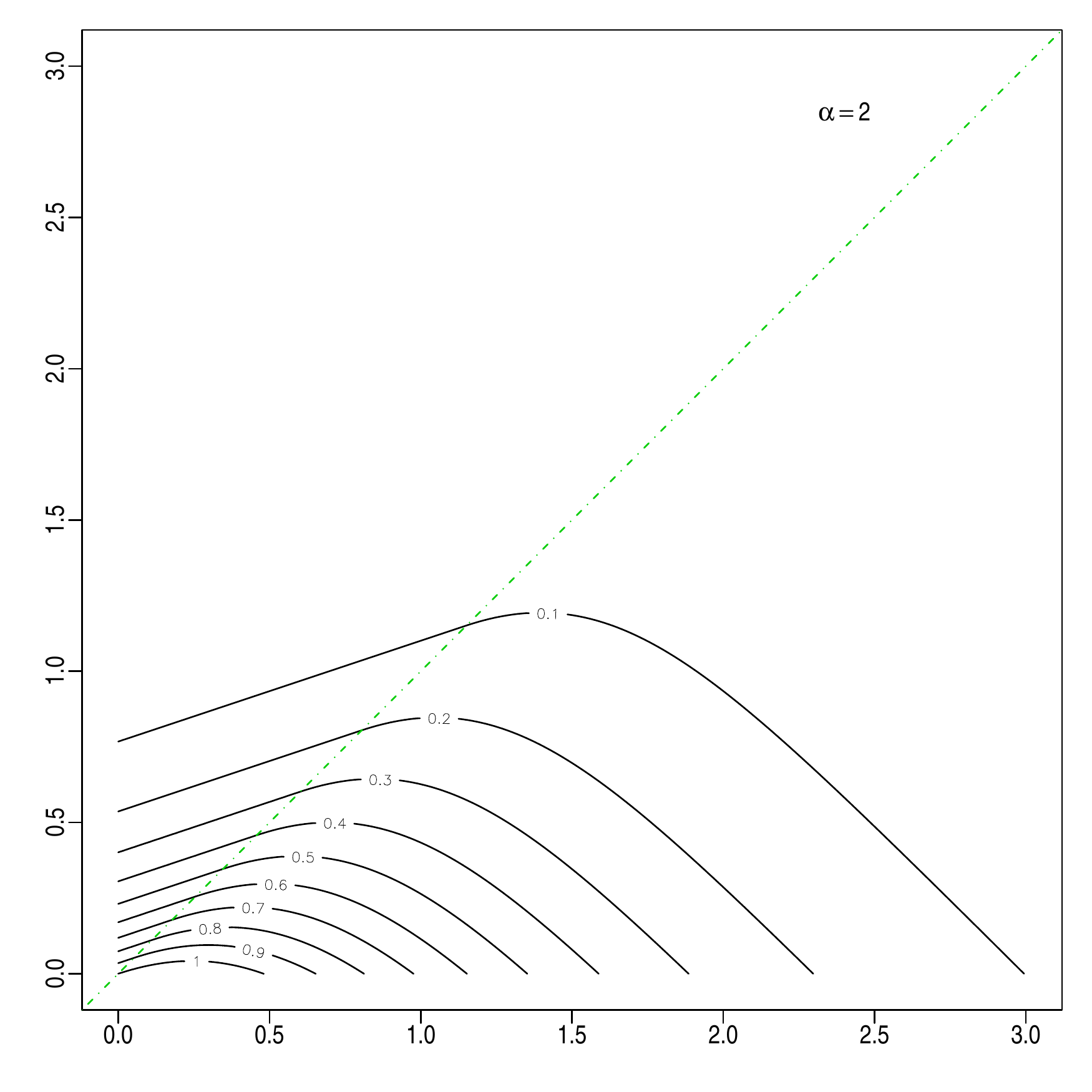} \hfil
   \includegraphics[width=0.48\textwidth]{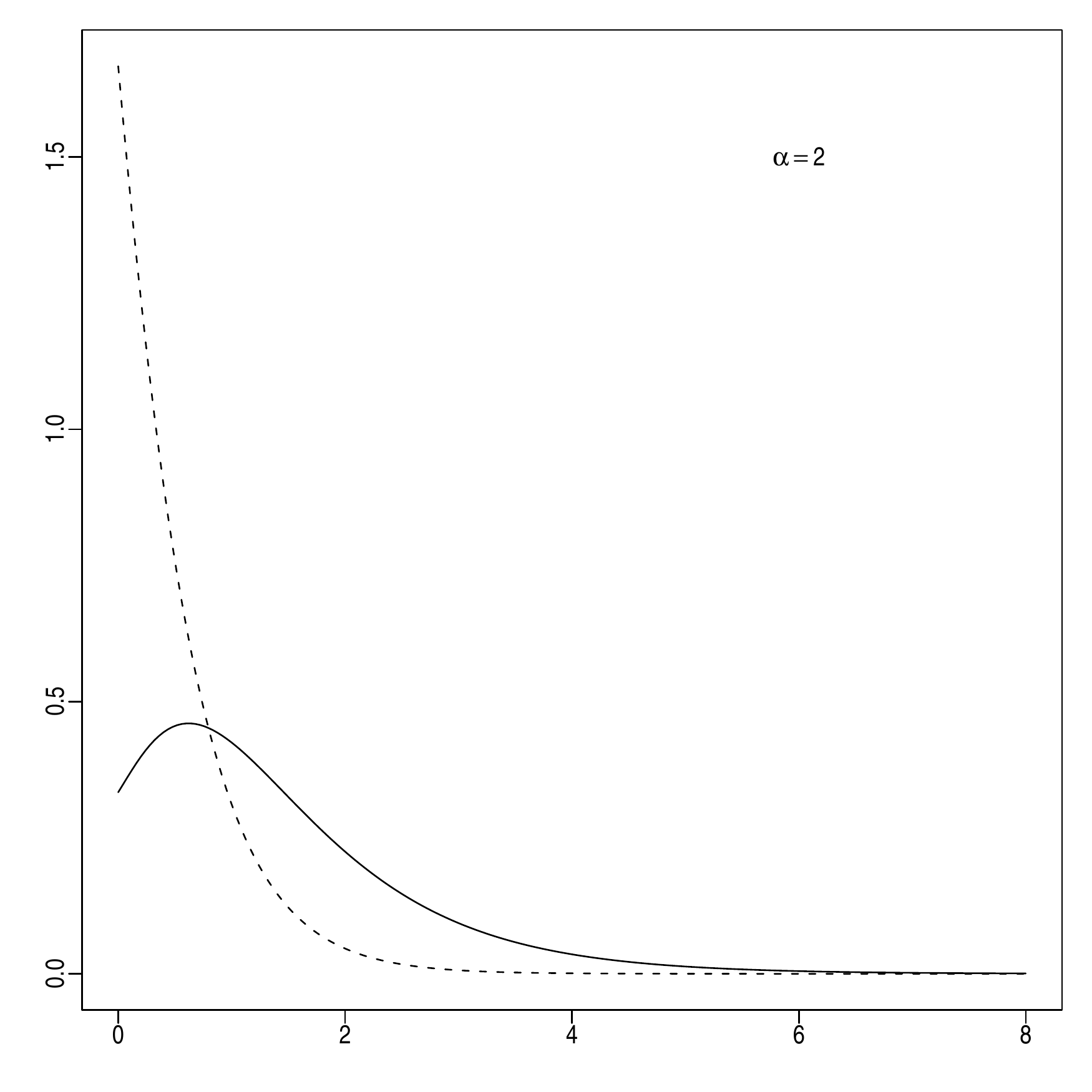}
   \caption{\sl Density function of $Z$ in the exponential-Laplace case when
     $w(y)=2(y_1-y_2)$ in the left panel, and marginal density of $Z_1$
     (continuous line)  and $Z_2$ (dashed line)  in the right panel }
   \label{fig:exp-laplace-pdf}
\end{figure}
Figure~\ref{fig:exp-laplace-pdf} displays the bivariate density function of
$Z$ and the marginal density of $Z_1$ and $Z_2$ when $\alpha=2$.

The moments of the marginal densities can be computed by direct integration,
which for $\alpha>0$ and any $r>0$ gives
\begin{eqnarray*}  
  \E{Z_1^r} &=& 
            \cases{
               \Gamma(r+1)\left(2 - \dfrac{r+4}{2^{r+2}}\right) 
                           &  if $\alpha=1$,  \cr
               \Gamma(r+1)\left(2 -\dfrac{1}{\alpha^2-1}  \left[
               \frac{\alpha^2}{2^r}- \frac{1}{(1+\alpha)^r}\right] \right) 
                           &  if $\alpha>0,\ \alpha\not=1$,
               } \\
   \E{Z_2^r} &=& 2\,\Gamma(r+1) - \E{Z_1^r}
\end{eqnarray*}  
while the moments for $\alpha<0$ are obtained by swapping the subscripts of
$Z_1$ and $Z_2$ in the above expressions and taking the absolute value of
$\alpha$. From  marginal moments up to second order and from the fact obtained
earlier that $\E{Z_1\,Z_2}=1$, we can compute the correlation.  Numerical
evaluation indicates that the correlation increases monotonically from 0 when
$\alpha=0$ to $1/\sqrt{5}\approx 0.447$ when $\alpha=\pm\infty$.

We close this section with some remarks on possible extensions.  The above
discussion has focused on the case with linear $w(y)=\alpha(y_1-y_2)$ but many
other options are possible, such as $w(y)=\alpha(y_1^2-y_2^2)$ or
$w(y)=\sin[\alpha(y_1-y_2)]$. Another direction is to adopt a base density
$f_0$ with dependent components, since our choice of using independent
components was only for mathematical simplicity.  
 The overview of bivariate exponential distributions 
presented in Section~47.2 of \citet{kotz:bala:john:2000} includes several
distributions which meet this requirement. Among them, two especially 
important proposals are the Gumbel and the Marshall--Olkin bivariate
exponential distributions, whose joint survival functions are
\[ 
   \exp\{- y_1 -y_2 - \lambda y_1\,y_2\} \,,\qquad
  \exp\{- y_1 -y_2 -\lambda \max(y_1,y_2)\}   
\] 
respectively, for $y_1,y_2 \ge 0$, when scale factors are not included; 
here $\lambda$ is a positive parameter, and $\lambda<1$ for the first case.

The  Marshall \& Olkin distribution presents the somewhat peculiar
feature of a positive probability mass assigned to the line $y_1=y_2$.
This situation is not handled directly by Proposition~\ref{th:lemma1999},
not even its extension stated at the end of Section\,\ref{s:intro},
because here $Y$ is not a continuous random variable.
The argument of Proposition~\ref{th:lemma1999} can be applied to $f_0$,
if this denotes the density function on the non-singularity set. 
The singularity set would not be affected, and a separate mechanism
can possibly be adopted to modify the probability distribution on this set.

\section{Final remarks}

%

We conclude with some remarks on cases with dimension $d>2$, when the base
function $f_0$ is centrally symmetric. To ensure that \ref{e:lemma1-pdf} is
a proper density we need that $w(Y)$ is symmetrically distributed around $0$
when $Y$ has density $f_0$. The case when $w(\cdot)$ is odd, which of course
includes linear functions, is covered by Proposition~\ref{th:lemma2003}, and
symmetry of $w(Y)$ holds in general.

For a non-odd $w$ which is a function of two components of $Y$ only, we can
still make use of Proposition~\ref{th:lemma1-w2} with a $d$-dimensional $f_0$,
by using the following argument. If $f_0$ is centrally symmetric, the same is
true for the distribution of any subset of its components. Hence a function
$w$ of the nominated components of the form required by
Proposition~\ref{th:f0*w2} produces a symmetric $w(Y)$.

We are then left with the case where $d>2$ and $w(Y)$ is a non-odd 
function of at least three components of $Y$. The characterization of the
set of such functions $w$ is a problem which does not appear amenable to
a simple solution, and it deserves a separate study, given also its
interest as a property of centrally symmetric distributions.

\subsection*{Acknowledgements}
I would like to thank Antonella Capitanio stimulating discussions on this
problem and Giuliana Regoli for several insightful remarks on a preliminary
version of the paper.


\end{document}